\documentclass[a4paper,11pt]{article}
\usepackage{jcappub}
\usepackage{amsmath}
\usepackage{amssymb}
\usepackage{graphicx}
\usepackage{color,ulem}

\title{Quantum Fisher Cosmology: Confronting Observations and the Trans-Planckian Problem}

 \author[a]{C\'esar G\'omez,} 
\emailAdd{cesar.gomez@uam.es}

\author[b,c]{Raul Jimenez} 
\emailAdd{raul.jimenez@icc.ub.edu}

\affiliation[a]{Instituto de F\'{i}sica Te\'orica UAM-CSIC, Universidad Aut\'onoma de Madrid, Cantoblanco, 28049 Madrid, Spain.}
\affiliation[b]{ICC, University of Barcelona, Marti i Franques 1, 08028 Barcelona, Spain.}
\affiliation[c]{ICREA, Pg. Lluis Companys 23, Barcelona, E-08010, Spain.}

\abstract{
    The aim of Quantum Fisher Cosmology is to use the quantum Fisher information about pure de Sitter states to derive model independent observational consequences of the existence of a primordial phase of the Universe of de Sitter accelerated expansion. These quantum features are encoded in a scale dependent quantum cosmological tilt that defines what we can call the de Sitter universality class. The experimental predictions are: i) A phase transition from red into blue tilt at a scale order $k= 1$ Mpc$^{-1}$ that naturally solves the cosmological {\it trans-Planckian problem}, ii) A spectral index for curvature fluctuations at CMB scales $k= 0.05$ Mpc$^{-1}$ equal to $0.0328$, iii) A tilt running at scale $k=0.002$ Mpc$^{-1}$ equal to $-0.0019$, iv) An enhancement of the amplitude of CMB peaks for extremely high multipoles ($l > 10^5$) that can provide a natural mechanism for primordial black hole formation as a source of dark matter, v) A lack of power at scales of $8$ Mpc with respect to the CMB scale that can explain the $\sigma_8$ tension. 
}

\begin{document}
\maketitle

\section{Introduction}
Popperian  methodology of science~\cite{Popper}, although too conservative for the followers of  more aggressive postmodernist approaches, some of them based on identifying the sociological roots of trusting, gives us a very good advise on how to evaluate, at least in a first approximation, the success of Natural Sciences. Popper's old fashion advise is to simply confront the theory with the experiment; defining a boorish hierarchy of trust. Natural Science should admit the risk to be falsified by experiments and impose severe constraints on using the resource of invoking ad hoc hypothesis. In many cases, the general issue of falsifiability is not easy to define properly and strongly depends on how many of the parameters of the theory are taken as accidental or anthropic. More fundamentally is the question of which among these parameters are essentially classical or quantum mechanical, a question that is at the origin of the multiverse discussion. 

In modern Cosmology a very fruitful working hypothesis is to assume the existence of an early period of accelerated expansion with a graceful exit \cite{Inf1,Inf2,Inf3}. This general assumption leads to a series of concrete quantitative predictions that can be falsified. However, the richness of different classical modelizations of the primordial period of accelerated expansion can reduce the value of the confrontation with experiments to be just a selection of a particular model among too many. The predictive power of the theory is enormously reduced if for any experimental output we can always find the corresponding model that will fit the data. 

This methodological problem acquires a new boost once we frame the problem in the more general context of quantum gravity, namely on what quantum mechanics says about the very possibility of accelerated expansion i.e. of the quantum mechanical consistency of de Sitter space time \footnote{This is a long story that goes beyond the modest aims of this note. In essence, the quantum inconsistency of eternal de Sitter \cite{Gia1,Gia2,Gia3} naturally leads to upper bounds on the number of e-folds and consequently implements, in a model independent way, an upper bound on the de Sitter quantum lifetime leading to a natural quantum mechanical graceful exit. For a Swampland approach to de Sitter consistency see \cite{Garg:2018reu,vafa2,vafa3}.}.

After all, the inflationary paradigm reduces to work with a de Sitter geometry equipped with an {\it external clock} that we model classically using the inflaton potential. Most of the quantum features of inflation reflect the quantum properties of the clock. Thinking of the clock degrees of freedom as a scalar spectator: a natural, and in principle, modest task, is to identify the universal, model independent, quantum features of these spectator modes. 

In pure de Sitter, the ground state for these modes is generically a very entangled squeezed state that evolves in conformal time. After some {\it coarse graining} we can evaluate the entanglement entropy as well as its dependence on time. Moreover, we can include the effect of quasi de Sitter slow roll by taking into account non linearities i.e. the interaction among these modes responsible for cosmological non gaussianities (see Ref.~\cite{Br} and references therein). However, besides this entropic information associated with the coarse grained density matrix, we can consider, for the spectator {\it pure state} in de Sitter, the amount of quantum Fisher information (see e.g. review in Ref.~\cite{Paris}) and its dependence on energy scale. This information, by contrast to the former von Neumann entanglement entropy, is non vanishing for the {\it pure state} and informs us on the intrinsic uncertainties in energy and time. 

\begin{figure}
    \centering
    \includegraphics[width=.9\columnwidth]{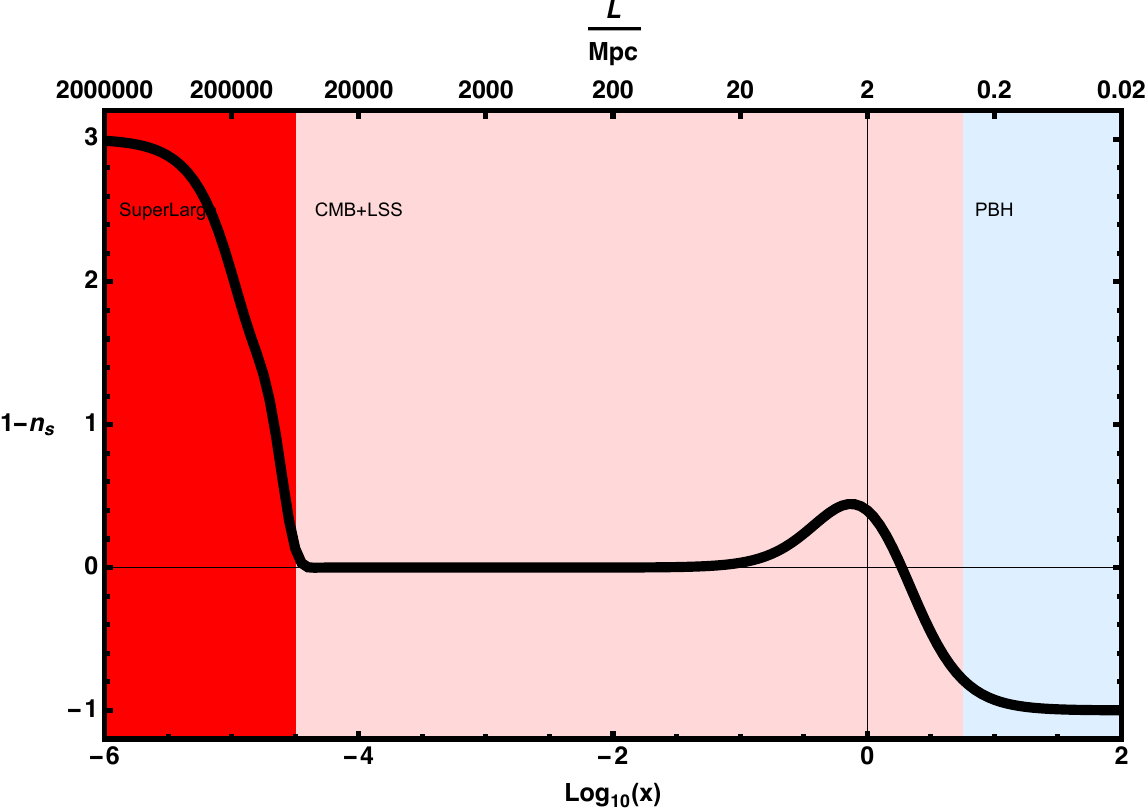}
    \caption{The cosmological tilt $1-n_s = \alpha_F/2$ as originally computed in Ref.~\cite{GJ6}. This tilt corresponds to sum over a number of entangled modes of the same order as the Gibbons-Hawking entropy. The argument $x$ is defined in (\ref{correspondence}). Also shown is the comoving length $L$ as computed in the text. The critical scale $k_{cr}$ is at about 1 Mpc.} 
       \label{fig:qtilt}
\end{figure}

In a series\footnote{Cosmological uses of quantum Fisher information in the context of relative entropy were initiated in \cite{GJ1,GJ2,GJ3}} of recent papers~\cite{GJ4,GJ5,GJ6} we have computed this quantum Fisher information. What we have found remarkable is that this quantum information scales, with respect to energy scale dilatations, in an {\it anomalous} way with a well defined and {\it scale dependent tilt}. This is a pure quantum effect that reflects the quantum phase dependence of the spectator quantum state in pure de Sitter \footnote{Although not very well known, quantum Fisher information is a very basic ingredient of de Sitter quantum dynamics. In its very bare bones it is simply the quantum Fisher information defined by the dependence on the external parameter of the family of de Sitter invariant vacua \cite{vacua1,vacua2} for the spectator modes. This quantum Fisher information encodes the intrinsic quantum uncertainties in de Sitter space-time.}. From now on the quantum cosmological tilt derived in \cite{GJ4,GJ5,GJ6} will be denoted $\alpha_F(x)$ with $x$ parameterizing the energy scale dependence. The explicit form of the quantum tilt is given in figure \ref{fig:qtilt}.

Motivated by this result, we have posed the {\it bold conjecture} that this quantum tilt contains full non-perturbative information on the physical cosmological tilt for scalar curvature fluctuations. In order to substantiate this conjecture, the first thing we do is to identify the spectral index by imposing, as consistency condition, the identity between the quantum energy uncertainty dictated by Fisher with the one expected in quasi de Sitter. This fixes a unique value of the spectral tilt at those scales where this identity is expected. Assuming that this scale is the CMB scale, we get a very precise prediction for the spectral tilt as well as a large set of predictions on the spectral tilt at different energy scales. An important output of this formalism is the prediction of two cosmological phases: red and blue tilted for soft and hard modes respectively (see Fig.~\ref{fig:qtilt}). The existence of the blue tilted phase for hard modes has important implications for the so called trans-Planckian problem of inflation \cite{BranMartin} that we will discuss briefly in this article. In essence, the blue tilted dynamics for hard modes defines an effective UV completion. 

Coming back to the former methodological discussion, this conjecture opens up the opportunity to associate with the existence of an early de Sitter phase of accelerated expansion a set of model independent predictions that both can be falsified by contrasting to the experiment but, by contrast to model dependent predictions, cannot admit potential corrections invoking some ad hoc changes of classical parameters. In this article we collect some of these predictions and we provide  numbers for the value of observables that our framework predicts implementing in that way an easy and clear path to falsify the model.

\section{Tilt and graceful exit}
One of the most robust predictions of inflation is the connection between the lack of scale invariance of the power spectrum and the duration of inflation. This connection, that goes back to the original Starobinsky model~\cite{Inf1}, was substantiated in Ref.~\cite{MCH} where the first identification of the cosmological tilt was done. Next we make some comments on this basic result.

Given a super horizon  energy scale $\lambda=k\eta H$ for $k$ comoving momentum and $\eta$ conformal time we define the associated number of e-foldings $N(\lambda)$ as
\begin{equation}
e^{-N(\lambda)} = \frac{\lambda}{H}
\end{equation}
If the end of inflation takes place at conformal time $\eta_{end}$ we define  ${\cal{N}}(k) \equiv N(k\eta_{end}H)$. Note that ${\cal{N}}(k)$ measures the number of e-folds between the time at which the mode of momentum $k$ exits the horizon and the end of inflation. In Ref.~\cite{MCH} the power spectrum for metric fluctuations at the moment they reenter the horizon is given by $Q^2(k)$ with $Q(k)$ defined by:
\begin{equation}\label{power}
Q(k) = \frac{3M}{M_P}(1+ \frac{1}{2}{\cal{N}}(k))
\end{equation}
with $M$ playing the role of the scalaron mass (as originally defined in the Starobinsky and Chibisov-Mukhanov papers).
Defining the tilt $n_s-1=\frac{d\ln Q^2}{d\ln(k)}$ we trivially get 
\begin{equation}\label{tilt}
(n_s-1)(k) = -\frac{2}{{\cal{N}}(k)} + \frac{4}{{\cal{N}}(k)^2} + ...
\end{equation}
For the observed tilt at $k=k_{\rm Planck18}$ the former expression leads to a neat prediction for the number of e-foldings ${\cal{N}}(k_{\rm Planck18})$. Reciprocally, if we know $\eta_{end}$ the former expression can be used to predict the value of the tilt. In inflationary models based on a slow roll inflaton potential, the value of $\eta_{end}$ is model dependent. It is determined by the point where the solutions to the Friedman equations for the given potential lead to violations of the slow roll conditions. The relevant prediction of Ref.~\cite{MCH} was the concrete relation between the tilt and the number of e-foldings, relation that turns out to be consistent with WMAP and Planck18 experiments. 

For ${\cal{N}}(k) >>1$ we can approximate
\begin{equation}
Q^2(k) = \frac{9M^2}{4M_P^2}{\cal{N}}^2(k)
\end{equation}
and $\frac{d\ln Q^2}{d\ln(k)} \propto \frac{2d \ln {\cal{N}}(k)}{d\ln(k)}$ leading to
\begin{equation}
{\cal{N}}(k) = {\cal{N}}(k_0) (\frac{k}{k_0})^{(n_s-1)(k_0)}
\end{equation}
as the {\it running} for ${\cal{N}}(k)$. Thus
\begin{equation}
Q^2(k) = \frac{9M^2}{4M_P^2}{\cal{N}}^2(k_0)(\frac{k}{k_0})^{2(n_s-1)(k_0)}
\end{equation}
Assuming that the amplitude at $k_0=k_{\rm Planck18}$ goes like $2 \times 10^{-9}$ for 60 e-foldings implies for the {\it scalaron} a mass $M\sim 10^{-6} M_P$. 

Using (\ref{tilt}) the running $\alpha \equiv \frac{d(n_s)}{d\ln k}$ evaluated at $k=k_{\rm Planck18}$ is given by
\begin{equation}\label{tilt1}
\alpha \sim -\frac{2}{{\cal{N}}^2(k_{\rm Planck18})}
\end{equation}

Thus Ref.~\cite{MCH} predicts a very low running of the order of $10^{-4}$.

\section{Quantum cosmological tilt prediction}
In Ref.~\cite{GJ6} the value of $1-n_s$ at CMB scales was identified in terms of the quantum tilt $\alpha_F(x)$ ( depicted in figure \ref{fig:qtilt}) as the solution to the consistency constraint
\begin{equation}\label{master}
\boxed{\alpha_F(3(1-n_s)) =2(1-n_s)}
\end{equation}
It is interesting to note that this relation is almost a {\it fixed point relation} that leads to
\begin{equation}
\boxed{1-n_s = 0.0328}
\end{equation}

We would like to stress the peculiarities of this prediction that only depends on requiring consistency between a general quasi de Sitter parametrization and the quantum Fisher description. The former number predicts what is the value of the tilt consistent with the simplest quasi de Sitter parametrization and, in this sense, it is fully model independent. In particular, this concrete number does not result from any assumption or input on the actual number of e-foldings.

Although the derivation of (\ref{master}) was presented in Ref.~\cite{GJ6}, we will summarize, for the reader convenience, the main steps to obtain this equation. 

\subsection{Brief review of Quantum Fisher Cosmology}
As stressed in the introduction, the key ingredient is the quantum Fisher information associated with the family of de Sitter invariant vacua describing scalar spectators in a pure de Sitter background. These pure states are sometimes denoted, in the literature, as $\alpha$-vacua. What the quantum Fisher information naturally defines is a metric on this set of states.  In essence it measures the quantum distinguishability of different $\alpha$ vacua (see Ref.~\cite{vacua1,vacua2} and also ~\cite{Bousso}  for a description of how these vacua are defined and related)\footnote{There exists an extensive literature on both the quantum consistency of $\alpha-$vacua as well as on the physical viability of using the Bunch Davis vacuum to define the quantum fluctuations describing the CMB spectrum of fluctuations (for some reference see \cite{Danielsson,Poly,Anderson,Graef}).Some of these problems are related with the trans-Planckian problem, that we will discuss in a separated section from a different point of view, and the computation of one loop effects on these vacua. We will surpass some of these well known difficulties focusing on a well defined quantity associated with the family of $\alpha$ vacua, namely the quantum Fisher information associated with this one parameter family of pure states. As stressed before, this quantum information naturally leads to a finite quantum variance for the parameters labeling these vacua.}. The main message of our work is that this quantum variance can account for the anomalous scale dependence of the cosmological power spectrum normally derived after adding a quasi de Sitter deformation. Moreover, this quantum information encodes the quantum variance of the $\alpha$ parameter. The parameter $\alpha$ can be associated with a natural energy scale defined as $k\eta H$ \footnote{More precisely $\alpha= ln tanh(r(\Lambda)) -2i\phi(\Lambda)$ with $r(\Lambda) 
=-sinh^{-1}(\frac{H}{2\Lambda})$ the standard squeezing parameter
and $\phi(\Lambda) =-\frac{\pi}{4} - \frac{1}{2} tan^{-1}(\frac{H}{2\Lambda})$
with $\Lambda = Hk|\eta|$ see \cite {GJ1}.}. 

The starting point of the quantum Fisher approach to Cosmology is to identify the scale transformations of this quantum Fisher information. In other words, we are interested in identifying how the information controlling the quantum variance of $\alpha$ depends on the energy scale at which we are working.  The main finding of reference \cite{GJ1} is that this scale transformation of the quantum Fisher is anomalous with a scale dependent tilt defined as $\alpha_F$. This is the tilt depicted in Fig.~\ref{fig:qtilt}. As discussed in Ref.~\cite{GJ3} this figure represents the numerical result obtained after evaluating the quantum Fisher information with an IR cutoff on the number of contributing entangled pairs. The sensitivity of the result on this IR cutoff was discussed in Ref.~\cite{GJ3} and we briefly review it below. 

\subsection{Sketch of the derivation of (\ref{master})}
The derivation of (\ref{master}) is based, as described in Ref.~\cite{GJ3}, in a three step argument. 

\begin{enumerate}
\item First of all you focus your attention on the quasi de Sitter modification of the effective frequency of spectator modes. This frequency is the one appearing in the basic Schrodinger-Chibisov-Mukhanov equation
\begin{equation}
\phi^{''}_k + (k^2 - \frac{a_{qdS}^{''}}{a_{qdS}})\phi_k =0
\end{equation}
In the simplest slow roll approximation we can use $\frac{a_{qdS}^{''}}{a_{qdS}} = \frac{\beta(\beta+1)}{\eta^2}$, where $^{'}$ denotes derivative with respect to $\eta$ and to define the slow roll parameter $\delta$ by $\beta = -2-\delta$ \footnote{This is the simplest {\it local} parametrization of quasi de Sitter where we ignore the scale dependence of $\epsilon$.}  As usual $a$ is the expansion factor of the FRLW metric. Now you extract the pure quasi de Sitter contribution to the oscillator energy that is given by

\begin{equation}
    \delta_{qdS} E^2 = \frac{3\delta+\delta^2}{a_{qdS}^2 \eta^2}
    \end{equation}

    Fixing a pivot scale $k_0$ and the qdS Hubble at that scale $H_0$ we get
$\delta_{qdS} E^2= (3\delta+\delta^2)H_0^2 (k_0\eta)^{2\delta}$. Note that this quantity depends on the a priori totally free phenomenological parameter $\delta$ defining the quasi de Sitter. 

\item Let us now focus on pure de Sitter. What we do at this point is to use the quantum variance defined by the quantum Fisher information to identify the corresponding quantum contribution to the oscillator energy. This is done by defining at first order (same order in $\hbar$ as the quasi de Sitter contribution) this contribution as

\begin{equation}
    \delta_F E^2 =\frac{k \delta_F(k)}{2a_{dS}^2}
\end{equation}
with $\delta_F(k)$ determined by the quantum Fisher information
\begin{equation}\label{Fis}
F = \frac{1}{\eta^2} (k\eta)^{\alpha_F(k\eta)}
\end{equation}
as $\delta_F(k) = F^{1/2}$ \footnote{Note that (\ref{Fis}) is defined fixing $k$ and deriving with respect to $\eta$. Thus this quantum Fisher measures the variance of $k^2$. In the linear approximation we are identifying $\delta_F(k)$ as the square root of $\delta_F(k^2)$}. 
Using the same pivot scale $k_0$ and $H_0$ we get
\begin{equation}\label{Fisher}
    \delta_F E^2 = H_0^2 (k_0\eta)^{\frac{1}{2}\alpha_F(k_0\eta)}(\frac{k_0\eta}{2})
\end{equation}
Note that this contribution depends on pure quantum de Sitter data with the key information encoded in the tilt $\alpha_F(k\eta)$. 

\item The final step requires to identify, locally, both contributions: namely the quasi de Sitter and the quantum Fisher. Since our task is to {\it predict} the value of the quasi de Sitter parameter $\delta$ we look for the value of $\delta$ such that $\delta_F E^2 = \delta_{qdS} E^2$. This equation identifies the value of the slow roll parameter that agrees with the quantum Fisher estimation.  This leads to the equation for $\delta$
\begin{equation}
    4\delta = \alpha_F( 6\delta +2\delta^2)
\end{equation}
with $\alpha_F$ as the model independent data. This for the slow roll parameterization $2\delta =1-n_s$ becomes equation(\ref{master}). Solving this equation we get the value of the slow roll parameter, as well as the energy scale at which the slow roll approximation agrees with the quantum Fisher result. 
\end{enumerate}

As a final comment we draw the attention of the reader to the already advertised almost fixed point form of equation (\ref{master}). Indeed, the trivial {\it fixed point} appears for the Harrison-Zeldovich scale invariant case $n_s=1$ corresponding to the argument of $\alpha_F$ equal zero i.e. in the limit of eternal de Sitter.

Next we will use the quantum tilt dependence on energy scale to make some further predictions. In order to do that we will chart the different energy scales using local {\it slow roll coordinates} leading to a scale dependence of the effective slow roll local parameters. 

\section{The meaning of the cosmological phases}
Once we introduce as data the value of the Hubble parameter $H$ at inflation and the current value $H_0$, we can divide the different modes with comoving momentum $k$ into {\it hard} and {\it soft} where roughly hard/soft modes are the ones with small/large value of ${\cal{N}}(k)$. Assuming instantaneous reheating ($rh$) we can consider as {\it soft} those modes satisfying
\begin{equation}\label{constraint}
e^{{\cal{N}}(k)} > \frac{H}{H_0}e^{-N_{rh}(k)}
\end{equation}
These are the modes for which the number of e-foldings accrued during inflation after horizon exit is larger than the total number of e-foldings $N_t$ defined by $e^{N_t}
=\frac{H}{H_0}$ minus the effective number of e-foldings  taking place between the end of reheating and the present time that we will denote $N_{rh}$ and that can be defined, up to the correction due to the number of species, as $e^{N_{rh}} = \frac{T_{rh}}{T_0}$ where $T_{rh}$ is the reheating temperature and $T_0$ is the temperature of the Universe today. For soft modes with low momentum $k\eta_{end} <<1$ the constraint (\ref{constraint}) is obviously satisfied. The complementary type of {\it hard} modes will be those with large value of $k$ for which 
\begin{equation}\label{constraint2}
e^{{\cal{N}}(k)} < \frac{H}{H_0}e^{-N_{rh}(k)}
\end{equation}
These are the modes that could give rise today to small size structures. Normally, it is assumed that the CMB mode $k_{\rm Planck18}$ are those for which
$e^{{\cal{N}}(k_{\rm Planck18})} \sim \frac{H}{H_0}e^{-N_{rh}(k_{\rm Planck18})}$. Thus in this qualitative decomposition of scales {\it hard/soft} modes are those with momentum $k$ larger/smaller than $k_{\rm Planck18}$. The two phases {\it red and blue} derived from the quantum Fisher tilt correspond qualitatively to {\it soft and hard} modes respectively. 

In order to make this classification a bit more precise we need to map the variable used as the argument of $\alpha_F$ with the comoving momentum $k$. This will be done mapping the solution of equation (\ref{master}) with the CMB data.

Hence denoting $x$ the argument of $\alpha_F$ we need to map $x$ to a physical comoving momentum $k$, let us say $x(k)$ and to fix this map by imposing $x(k_{\rm Planck18}) \sim 0.1$. This can be done in a linear approximation in order to extract from the quantum cosmological tilt concrete predictions about other energy scales. However, and using the former qualitative characterization of hard and soft modes, we can define
\begin{equation}\label{correspondence}
x(k) = \frac{k\eta_{end}}{e^{-A}}
\end{equation}
where $A$ is
\begin{equation}
e^{-A} \equiv \frac{H_0}{H} \frac{10 T_{rh}}{T_0}
\end{equation}
The factor 10, that is just a phenomenological fit, designed to make $x(k_{\rm Planck18}) \sim 0.1$, can be reabsorbed in a redefinition of the reheating temperature. It is important to stress that although in our formalism the tilt $\alpha_F$ is totally independent of the details of the post-inflationary phase the specific map, defined in (\ref{correspondence}), into {\it current} experimental data, depends on the characteristics of the post inflationary phase in particular on the reheating temperature as well on the current value of the Hubble parameter. We fit these data locally in a linear approximation around the CMB scale.

Using this phenomenological parametrization we can identify the two {\it cosmological phases} predicted by the quantum cosmological tilt namely:
\begin{itemize}
\item Phase I ( red) for $\frac{k\eta_{end}}{e^{-A}} < 1$
\item Phase II (blue) for $\frac{k\eta_{end}}{e^{-A}} > 1$
\end{itemize}
Defining a time scale $N_F$ by 
\begin{equation}
\boxed{N_F \sim \ln(\frac{H}{H_0}\frac{T_0}{10 T_{rh}})}
\end{equation}

the two phases are characterized by modes for which ${\cal{N}}(k) > N_F$ (red modes) and those for which ${\cal{N}}(k)< N_F$ (blue modes). In other words, the red Phase I corresponds to modes for which the time, measured in e-foldings, between horizon exit and the end of inflation is bigger than the scale $N_F$ and the other way around for the blue Phase II.

\begin{figure}
    \centering
    \includegraphics[width=0.8\columnwidth]{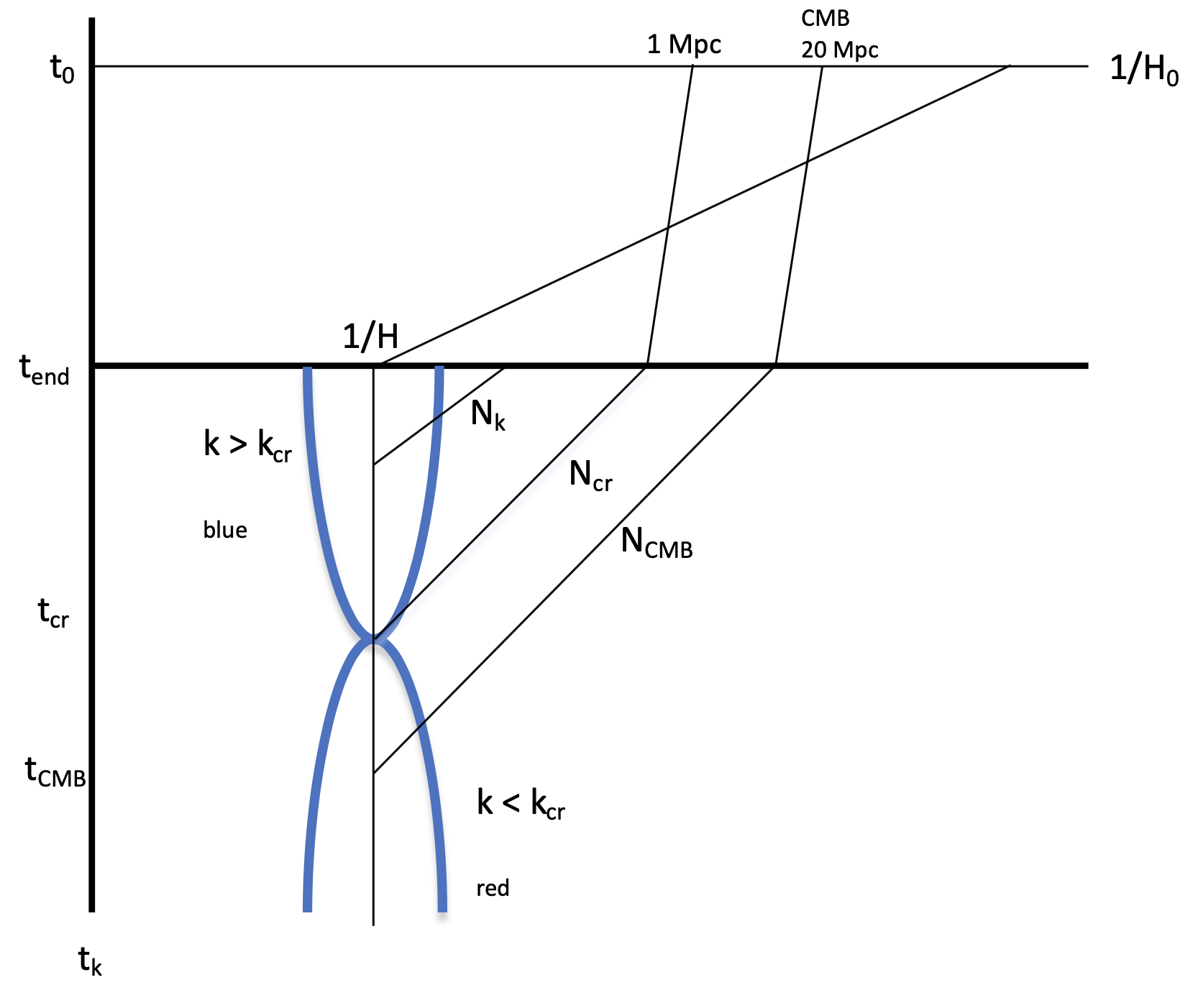}
    \caption{Graphical depiction of modes leaving the horizon for different times. $t_k$ is the time at which the mode leaves the horizon. $N_k$ is the number of e-folds for a mode $k$ between exit and end of inflation. $k_{cr}$ denotes the critical scale of $\sim 1 $ Mpc.}
    \label{fig:fig1}
\end{figure}

In order to get a more intuitive picture of the phases, we can use the space-time diagram in Fig.~\ref{fig:fig1}. For a generic comoving mode $k$ we define $t(k)$ as the time at which this mode exits the horizon. From $t(k)$ to $t_{\rm end}$ representing the end of inflation the physical wavelength of this mode is enhanced by a factor $e^{{\cal{N}}(k)}$. During the post inflationary phase this wavelength is again expanded by a factor $e^{N_{rh}(k)}$, in the notation used above, ending at some current scale represented in the upper horizontal line. Now, {\it soft} modes with small $k$ correspond to early times $t(k)$ and larger values of ${{\cal{N}}(k)}$. The CMB scale $k_{\rm CMB}$ is the one we normally associate with ${{\cal{N}}(k)} \sim 60$ representing a current length scale of the order $20$ Mpc. {\it Hard} modes with $k$ ten times larger than $k_{CMB}$ define current smaller length scales and they correspond to smaller values of ${{\cal{N}}(k)}$. 

We can identify with the phase transition a $k_{cr}$ such that modes with $k>k_{cr}$ are {\it blue} and modes with $k<k_{cr}$ are red. This $k_{cr}$ also defines a $t_{cr}$ as the time at which modes with this momentum exit the horizon. The current length scale associated with $k_{cr}$ is of the order of 1 Mpc. Indeed from (\ref{correspondence}) the critical point  at which we pass from red into blue phase corresponds to 
\begin{equation}
\boxed{k\sim 20 \times k_{\rm Planck18} \sim 1 {\rm Mpc}^{-1}}
\end{equation}
for $k_{\rm Planck18}= 0.05 {\rm Mpc}^{-1}$ \footnote{Note that in order to estimate $k_{cr}$ we first fix the correspondence at $x(k_{\rm CMB})=0.1$ and next we look for the effective $k$ at wich $\alpha_F$ changes from red into blue.}. 

Any {\it upper bound} ${{\cal{N}}_{max}}$ on ${{\cal{N}}(k)}$ defines a natural IR scale $k_{IR}$ by ${{\cal{N}}(k_{IR})} = {{\cal{N}}_{max}} $. Reciprocally, the natural UV scale $k_{UV}$ can be defined by the condition
${{\cal{N}}(k_{UV})} = 0$. This is the larger value of $k$ that exits the horizon before the inflation ends. It is the mode that exits the horizon at precisely the end of inflation. 

The interpretation and meaning of an upper bound on the number of e-foldings has been recently the subject of several analysis. In particular the TCC (Trans-Planckian-Censorship-Conjecture) \cite{vafa} requires that $k_{UV} \leq M_P$. In this case and since, by definition, $k_{UV}= \frac{1}{\eta_{end}}$, we get for $\eta_{end} = \frac{1}{H} e^{-{{\cal{N}}_{max}}}$ that the condition $k_{UV}\leq M_P$ leads to the TCC bound 
\begin{equation}\label{TCC}
e^{-{{\cal{N}}_{max}}} \leq \frac{M_P}{H}
\end{equation}
 The meaning of the TCC conjecture is to restrict the harder mode $k$ that can exit the horizon before the end of inflation to be the Planck scale, solving in this way the ultraplanckian problem. This condition however imposes superficially severe restrictions on slow roll inflationary models. For instance for ${{\cal{N}}}_{max} \sim 60-70$ the TCC will implies a very low value of $H$ at inflation. Moreover in order to accommodate slow roll with this form of TCC requires a very unnatural hierarchy between the slow roll parameters $\epsilon$ and $\eta$ with $\eta$ being almost 30 orders of magnitude larger than $\epsilon$ \cite{BranVafa} (see also \cite{Mizuno:2019bxy,Seleim:2020eij}).

Although we don't want to enter here into the deep discussion about Wilsonian decoupling in an expanding Universe (see the critical comments in \cite{Burges}) we would like to point out to the change of perspective on the whole discussion induced by the existence of a blue tilted phase for hard modes. 

\section{The physics of the blue tilted regime and the trans-Planckian problem}
Naively the TCC is motivated by the following tension. If by some argument we discover a maximal number of e-foldings ${{\cal{N}}}_{max}$ for a given $H$ we can have created a trans-planckian problem if, by formally reverting the time, we discover that the wave length of the primordial mode that after ${{\cal{N}}}_{max}$ e-foldings becomes $\frac{1}{H}$ i.e. exit the horizon, is ultraplanckian.  Our target in this brief section is to show that the existence of the blue tilted phase for hard modes naturally solves the trans-Planckian problem.

\begin{figure}
    \centering
    \includegraphics[width=0.8\columnwidth]{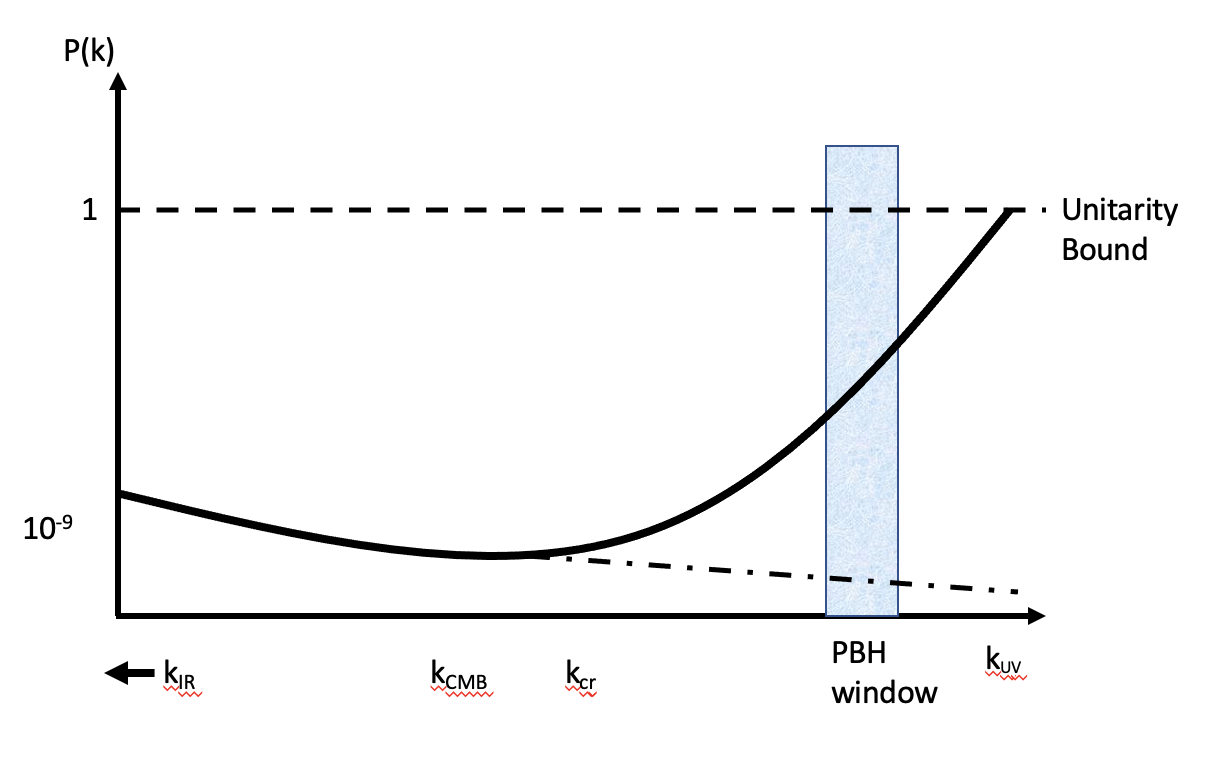}
    \caption{The two phases and the trans-Planckian problem. The solid line shows the power spectrum for soft modes (red tilted) while the dashes-dotted line the one for hard modes (blue tilted).}
    \label{fig:fig2}
\end{figure}

The natural physical way to impose an IR limit on the value of the smallest comoving momentum $k_{IR}$ that can exit the horizon is observing that, when we move into the IR, since we are in the red tilted phase, the power spectrum is increasing as $(\frac{k_0}{k})^{1-n_s}$ with $(1-n_s)$ positive. This is a local behavior around a pivot scale of the order of $k_0=k_{\rm CMB}$. In order to extrapolate this behavior for small $k$ far away from the pivot CMB scale we need to assume that $(1-n_s)$ is not changing too much when we move into the IR. From the quantum tilt (see Fig.~\ref{fig:qtilt}) we observe that this is the case and that $(1-n_s)$ is indeed remaining positive although decreasing in the IR. If we ignore this small running the limit on the value of $k_{IR}$ can be naturally defined by the {\it unitarity} condition that the amplitude should be smaller than one. Very naively this will lead to a limit on $k_{IR}$ of the order
\begin{equation}
(\frac{k_0}{k_{IR}})^{(1-n_s)} \sim 10^9
\end{equation}
that as can be easily seen can be extraordinarily small for $(1-n_s) \sim 10^{-2}$. For such $k_{IR}$ the number of associated e-foldings ${{\cal{N}}}_{k_{IR}}$ can be much larger than the ${{\cal{N}}}_{max}$ determined by TCC. Note that this argument strongly depends on keeping the red tilt small in the IR.

 To check the meaning of the TCC we need an {\it independent} estimate of the maximal value of the comoving momentum $k_{UV}$ that can exit the horizon before the end of inflation. How to estimate this value without assuming TCC? The answer is simple if again we use the amplitude of the corresponding power spectrum. If for $k$ larger than the pivot scale the tilt {\it continues to be red} them the amplitude will diminish when we move into the UV. This precludes to impose any unitarity bound on how hard can be $k_{UV}$ and consequently either we deal with an ultraplanckian problem or we decide to accept the TCC limit. However, if for hard modes the tilt becomes blue, then the whole situation changes dramatically. Indeed such a case means that whenever we consider the UV regime with $k>k_{cr}$ the blue tilt induces an {\it increasing} amplitude that goes like
\begin{equation}
(\frac{k}{k_{cr}})^{n_s-1}
\end{equation}
with $n_s-1$ positive in the blue tilted region. The quantum Fisher tilt provides enough information to get a rough estimate of the maximal value $k_{UV}$, namely
\begin{equation}\label{boundUV}
(\frac{k_{UV}}{k_{cr}}) \sim 10^9
\end{equation}
where we take for simplicity the limit of the blue tilt $n_s=2$. For a cartoon of the physical picture we are obtaining see Fig.~\ref{fig:fig2}. In this figure we observe that the amplitude of the power spectrum grows, although extremely slowly, in the IR regime that is almost scale invariant. In the UV regime the amplitude grows faster as indicated by the quantum tilt. At the critical point $k_{cr}$ we reach an effective minimum. In an extremely speculative mood, and by analogy with QCD like theories, we could try to identify the length associated with the critical point, that is of the order of 1 Mpc, as a sort of {\it dynamically generated cosmological scale}. Note also that the $k_{UV}$ defined by (\ref{boundUV}) is of the order $e^{N_{rh}}10^{9}{\rm Mpc}^{-1}$ that for realistic values of $N_{rh}$ should be $\leq M_P$. From this point of view the Planckian cosmological cutoff i.e. the trans-Planckian censorship induced by the existence of the blue tilted phase, becomes \footnote{For realistic models of inflation with instantaneous reheating the so defined bound on the UV scale is close to Planck scale, up to details that depend on the number of species created after reheating and on the reheating temperature. }
\begin{equation}
   \boxed{e^{N_{rh}}10^{9}\frac{1}{{\rm Mpc}}\leq M_P} 
\end{equation}
that should be compared to the TCC conjecture (\ref{TCC}).
 In the figure we superimpose the behavior in the case of a pure red tilt that clearly makes manifest the ultra-planckian problem \footnote{It could be of some intrest to read the curve in figure 2 in the spirit of Page's curve \cite{page}. As a very superficial comment note that the quantum Fisher information starts to grow significantly precisely at $k=k_{cr}$.} . 

In words: {\it the existence of a blue tilted phase naturally nullifies the trans-Planckian problem in inflation.} The reason being that the blue tilt naturally imposes a unitarity bound on the larger value of $k_{UV}$ in a way that is not constraining ${{\cal{N}}}_{k_{IR}}$. This last number depends on the IR behavior of the quantum tilt and can be in principle as large as the one predicted by general quantum arguments as the quantum breaking time \cite{Gia3}. Moreover the model independent blue tilted dynamics for hard modes, predicted by the quantum tilt, effectively defines a {\it UV completion} that naturally avoids
the ultraplanckian problem.

As a final comment let us briefly come back to the discussion on the IR scale. As pointed out once we assume a small red tilt in the IR the bound on the maximal number of e-foldings coming from the identification of $k_{IR}$ can be much larger than the one predicted by TCC. However at this point we should recall some technical aspects of our computation of the quantum tilt in the very IR region that were discussed in \cite{GJ3}. In figure (\ref{fig:qtilt}) it appears a very red tilted regime in the ultra soft region. This increase in the red tilt can changes dramatically the behavior of the power spectrum in the IR modifying the expected value of $k_{IR}$. As discussed in \cite{GJ3} the numerical analysis used to derive the tilt strongly depends in the IR on the number of entangled pairs contributing to the quantum Fisher information. On the contrary the behavior of the quantum tilt in the UV is very independent on this cutoff. As we discuss in section 7 a natural bound on the number of contributing entangled pairs can be derived using the standard de Sitter entropy. Using this prescription the number of e-foldings associated with the corresponding $k_{IR}$ is still much larger than the one predicted by the TCC prescription.

\subsection{Primordial black holes}

As it was originally observed in \cite{ZelNov,CarrHawking,Hawking} primordial black holes can be created if the amplitude of the primordial fluctuation is large enough to lead to gravitational collapse once we enter into the post inflationary radiation dominated phase. This is impossible if the cosmological tilt remains red to all scales but it is possible in the blue tilted phase for some modes with $k>k_{cr}$. To identify the modes that can lead to the formation of PBH is a question that requires to introduce some astrophysical constraints on the value of the $\beta$ parameter defined as the portion of matter that can collapse into black holes i.e.
$\beta \sim \frac{\Omega_{\rm BH}}{\Omega_{\rm matter}}$. Assuming gaussianity this quantity can be related to the power spectrum and leads to a window for PBH for scales
\begin{equation}
(\frac{k}{k_{cr}}) \sim 10^7
\end{equation}
As can be seen in Figure 2 this region, assuming that the blue tilt extends to larger UV scales is smaller than the naive estimate of $k_{UV}$ by two orders of magnitude. Of course this is very qualitative but seems to indicate as a natural possibility that the $k_{UV}$ should be fixed by the threshold of not too big primordial black hole formation. However this comment depends on details of the post inflationary phase about which we have not made any assumption. To study in more detail this interesting possibility, that requires astrophysical input, goes however beyond the scope of this note.

\section{Tilt running}
As already pointed out standard inflationary models lead to a very small tilt running of the order of $\frac{1}{(1-n_s)^2}$. This prediction, although consistent with Planck results, is a bit too small. Next we present the prediction on the tilt running derived from the quantum Fisher approach.
The Planck satellite has measured the running of the tilt, i.e.

\begin{equation}
n(k) = 1-n_s + (1/2) (dn_s/d \ln k) \ln (k/k_0)
\end{equation}

When allowing for running, the Planck18 data give the following constraint, $ dn_s/d \ln k =  -0.0041 \pm 0.0067$ when including also information about baryonic acoustic oscillations from the BOSS survey. This logarithmic derivative of $n_s$ is computed at a scale of $k=0.002$ Mpc$^{-1}$, i.e. 25 times larger than the scale at which $n_s$ without running is computed.

We can do the same in our model.  To do this we compute it at $|k_F \eta_F| = 0.004$, which is 25 times larger than our slow-roll solution. In our case this leads to:

\begin{equation}
\boxed{ dn_s/d \ln k = -0.001911}
\end{equation}

This value is compatible with the Planck18 one but typically a factor ten larger than predictions from single field slow roll models (see (\ref{tilt1})). This makes it possible to falsify this prediction with future observations of CMB spectra distortions like Pixie/LiteBIRD.

\begin{figure}
    \centering
    \includegraphics{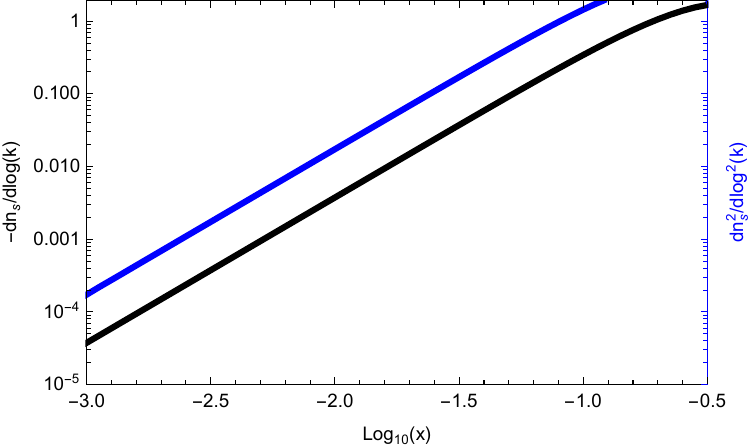}
    \caption{Running (black) and running of the running (blue) of the spectral index as a function of scale ($x$ is defined in (\ref{correspondence})).}
    \label{fig:running}
\end{figure}

\section{An Entropic approach to the end of inflation}
The reader should note that since we are not introducing any quasi de Sitter input we can only use quantum mechanical arguments to put an upper limit on the duration of inflation. The simplest upper bound can be defined using the IR cutoff (see Ref.~\cite{GJ5}) on the number $n$ in the sum over entangled pairs used for the numerical computation of the quantum cosmological tilt. Approximating the entanglement entropy of the $n$ pairs to be $O(n)$ i.e. each pair contributing as $\ln (2)$, the natural bound will be $n<N_{GH}$. This implies, numerically, a scale $x \sim \frac{1}{\sqrt{N_{GH}}}$. 

In order to estimate this value we will use the following very qualitative argument. Using (\ref{power}) it is natural to define an upper bound on the number of e-foldings given by the condition $Q=1$  that leads to \footnote{ It is important to stress that this estimation on the life time of the de Sitter primordial phase is different to the one originally used in \cite{MCH} associated with $P(k)$. }.
\begin{equation}\label{e-folds}
{\cal{N}}_{max} = \frac{2M_P}{3M} \sim 10^6
\end{equation}
If we now compare with the {\it compositeness bound~\footnote{ This bound was originally obtained on the basis of taking into account quantum effects for a purely gravitational coherent state model of de Sitter geometry. We use this constraint as simply a way to estimate, in the frame of \cite{MCH} a lower bound on $H$, for $H$ the inflationary scale.}} \cite{Gia1} 
${\cal{N}}_{max} \sim N_{GH}^{2/3}$ leads to the reasonable prediction on the value of $H$ at inflation of the order
\begin{equation}\label{hubble}
\boxed{H\sim 10^{-9/2} M_P}
\end{equation}

Recalling that $V= \frac{M_P^2}{4 \pi} H^2$ and that $V^{1/4} \sim r^{1/4} (2 \times 10^{16} {\rm GeV})$, we get $r \sim 0.006$ in agreement with our previous prediction~\cite{GJ2}.

Using this value of $H$ we get $x\sim 10^{-4.5}$. This is the case represented in figure\ref{fig:qtilt}. In linear approximation and using as pivot the CMB result at $x=0.1$ we get a natural estimate of the associated length scale $O(2 \times 10^{9/2})$ Mpc which is $\sim 10$ times larger than the size of the current observable Universe ($8$ Gpc) as well as on the associated number of e-foldings $O(6.10^5)$ in nice qualitative agreement with (\ref{e-folds}).

 If we are terribly optimistic, we can interpret this result as implying that the natural entropic bound, on the number of entangled modes contributing to the quantum Fisher, leads to a limit IR scale that is larger than the size of the visible Universe but not absurdly large.

At this point we will allow us a small digression on some natural implications of the former estimates.

If we consider that we need, at least, one galaxy (which typically is hosted in a dark halo of comoving size $> 0.1$ Mpc) to call something a Universe, then in log space, our current Universe is at a typical point in this space (3 out of 5) as it should be if we were to choose a Jeffrey's prior on our likelihood: there is nothing special about us. In other words, from a Bayesian prior point of view, if asked how big the Universe is, the answer should have been that since we are not special (Copernican principle) a Jeffrey's prior should have indicated that in log (size) we are in the middle.
The inflated patch of the Universe is thus $\sim 10$ times larger than our current visible universe. 

One strong prediction then, is that a measurement of the global curvature $\Omega_k$ must deviate at the level of 10\% from the current level of fluctuations $10^{-5}$ as large scale inhomogeneities are at scales $10$ times larger than our current horizon~\cite{Raca}. Because these large scale inhomogeneities have not been inflated away, the expectation is that they are $O(1)$ instead $O(10^{-5})$. A naive back-of-the-envelope estimate is that if we were able to measure curvature today at a suficiently precise level (as has been claimed e.g. in Ref.~\cite{Raca}), these $O(1)$ inhomogeneities should contribute order 10-20\% to our measurement of the global curvature, which should be at the $10^{-5}$ level if inflation is correct. The current bound, assuming the LCDM model by Planck18 is $\Omega_k = 0.001 \pm 0.004$ (at 95\% confidence), still two orders of magnitude away from the required precision.

Moreover recall that inflation in itself imposes a lower limit to the value of the global curvature of $\Omega_K > 2 \times 10^{-5}$. In our case, we are predicting that the size of the inflated patch of the Universe is $\sim 10^2$ Gpc. Beyond this patch, we should expect that inhomogeneities are of $O(1)$ as they have not been inflated away. If we take this as the radius of curvature $R_k$ of the Universe, we can compute the value of the curvature as $|\Omega_K| = (a_0 \frac{H_0}{c} R_K)^{-2}$; in our case $|\Omega_K| \sim 0.001$, which is just a factor 4 lower than the current Planck18 limit within the LCDM model. A detection of curvature above the cosmic variance limit of $\Omega_K > 2 \times 10^{-5}$ would be consistent with our predictions.

\section{Confronting with the $\sigma_8$ tension and enhanced high CMB multipoles.}

The above mapping of the Fisher argument $x$ into comoving scales allows us to compare our predictions with observations, as the quantum Fisher tilt provides an initial power spectrum as a function of comoving scale $1/k$ Mpc as depicted in Fig.~\ref{fig:qtilt}. 

We have identified $x =0.1$ as the scale at which the CMB has measured the tilt, i.e. where the slow roll approximation works.  In the case of the Planck18 satellite~\cite{Planck18}  $k = 0.05$ Mpc$^{-1}$. Thus we observe that for scales $1/k$ such that  $8 > \frac{1/k}{\rm Mpc} > 1$ the tilt is predicted to be redder than the one corresponding to the slow-roll regime and at much larger scales. This means the following: the usual way to proceed is to measure the initial power spectrum at CMB scales, and assuming it a power law with a fix index (in this case $n_s-1 = -0.0328$), extrapolate the value of the fluctuations at smaller scales. If the model is correct, when comparing this extrapolated value with other independent indicators, such as low redshift probes of the power spectrum or small scale CMB experiments, it should agree if it was the same power law for all scales. There are recent observational indications that this might not be the case.

On the observational side, the value for $1-n_s$ as inferred from the  low redshift ($z \sim 1$) measurement of the level of fluctuations by weak lensing surveys, specially KiDS~\cite{KiDS}, shows a lack of power with respect to the one inferred by assuming the value of $1-n_s$ derived from the CMB at the Planck18 satellite~\cite{Planck18}  $k = 0.05$ Mpc$^{-1}$ scale. This is also the case for the level of fluctuations measured directly at the CMB by small scale CMB experiments like the SPT-3G telescope~\cite{SPT}, i.e. when measuring directly the amplitude of the fluctuations at scales of $\sim 8$ Mpc. This lack of power seems to fit well with the predictions of the Quantum Fisher cosmological tilt. Clearly more data are needed to explore this full region as our prediction is very definitive. However, we can already explore in a  more quantitative manner the region around $\sim 8$ Mpc. and see how our predictions match the observations.

\begin{figure}
    \centering
    \includegraphics[width=0.8\columnwidth]{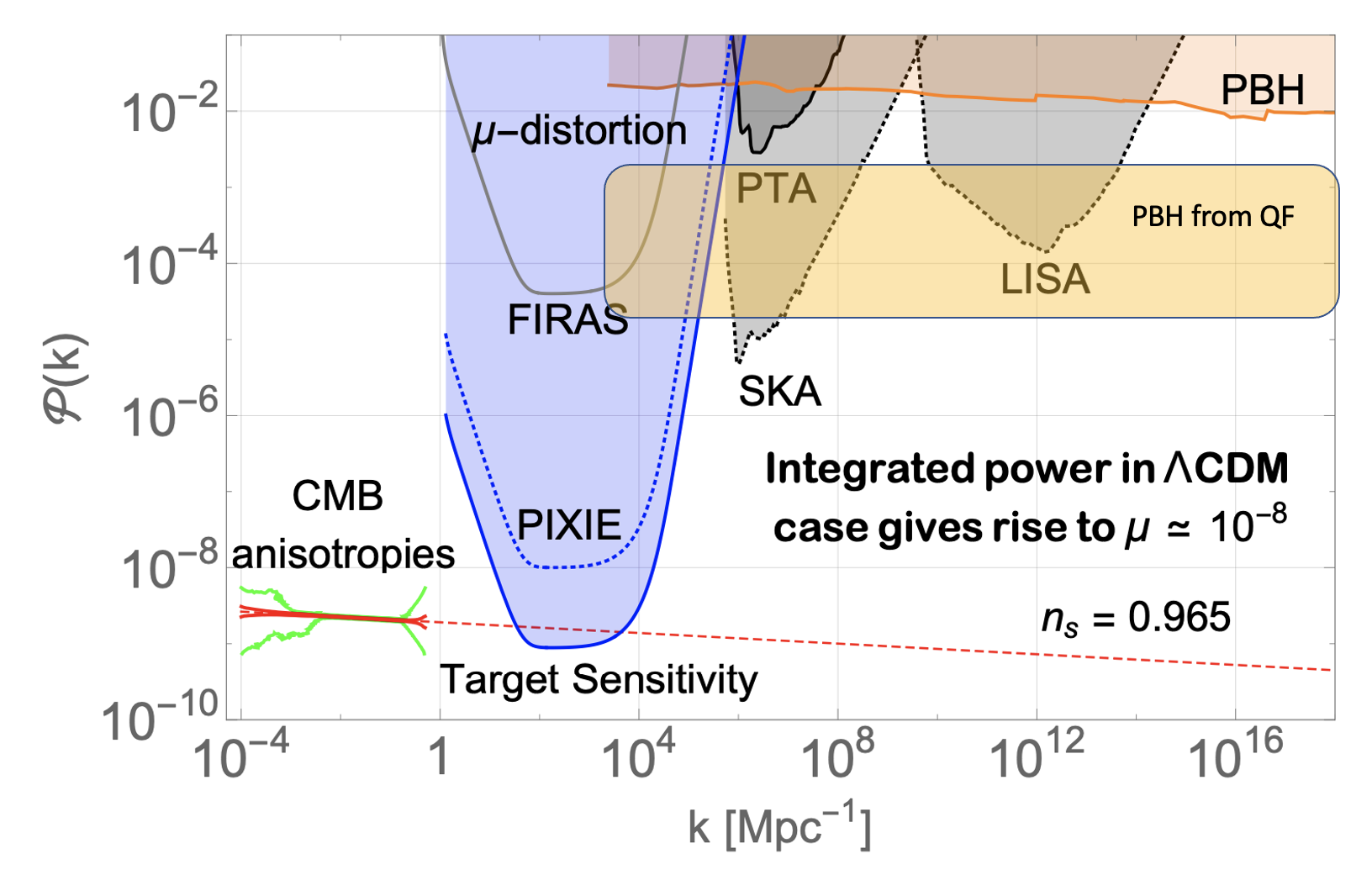}
    \caption{Figure adapted from Ref.~\cite{Chluba} to show the location of our predicted PBH.}
    \label{fig:pbhcmb}
\end{figure}

For comoving lagrangian scales $8 > \frac{1/k}{\rm Mpc} > 1$ we predict that the value of $1-n_s$ should be larger than at scales of $20$ Mpc, which is where Planck18 measured the tilt. Therefore, there should be a lack of power with respect to the CMB inferred one given by the difference in a power-law power spectrum when adopting the CMB value $1-n_s=0.0328$ or the value from Fig.~\ref{fig:qtilt} corresponding at these scales ($\log_{10} (x) \sim -0.5$). In more detail:

\begin{equation}
\Delta P(k)=\frac{P(k)_{\rm CMB}}{P(k)_{\rm \sim 8 Mpc}} = \frac{(k/k_0)^{(n_s-1)_{k \eta = 0.1}}}{(k/k_0)^{(n_s-1)_{k \eta \sim 0.3}}}    
\end{equation}

where we have used that the difference in scales between Planck18 CMB and the weak lensing surveys at $\sim 8$ Mpc is about a factor 3. Now, from Fig.~\ref{fig:qtilt} at the corresponding scale of $k \eta \sim 0.3$ we see that $1-n_s \sim 0.15$, so 

\begin{equation}
\Delta P(k) = \Delta(k)^{(0.0328 - 0.15)}
\end{equation}

where $\Delta(k) = 3$ as measured from the CMB scale of $\log_{10} (|k \eta|) \sim -1.0$ (to probe scales of 8 Mpc which is where KiDS and SPT report their measurements). This results in $\Delta P(k) = 0.85$. On the other hand the observed ratio $\Delta P(k) [S_8 ({\rm KiDS})/S_8 ({\rm Planck18})] =  0.9 \pm 0.03 $. This is in good agreement with our prediction that at small scales the universe whould be less clumpy than the one at CMB scales.

Obviously, more data are needed to constrain the scales $8 < 1/k < 1$ Mpc and confirm if the lack of power with respect to the CMB matches our prediction as seems to be the case from the KiDS weak lensing and small scale CMB data.

As we have shown in Ref.~\cite{GJ5} and discussed above, quantum fisher cosmology provides a natural transition from red to blue primordial spectrum at sufficiently small scales to generate fluctuations that result in primordial black holes. This happens for comoving scales $< 0.01$ Mpc. This very blue power spectrum (see Fig.~\ref{fig:qtilt}) will result in a large enhancement of the CMB temperature power spectrum peaks at very small scales. This will be measured by CMB spectral distortion experiments like LiteBird/Pixie (see Fig.~\ref{fig:pbhcmb} adapted from Ref.~\cite{Chluba}). This blue index results in an enhancement of the CMB Doppler peaks that goes like $l^{(n_s-1)}$. This enhancement will be better measured in the spectral distortion of the CMB temperature spectrum as the power spectrum is contaminated by secondary effects (point sources, thermal SZ, etc...) at these small scales ($l > 10^5$).

It is worth also noting that at comoving scales smaller than $1$ Mpc the tilt turns bluer. This should result in an enhancement of the number of halos of sub Milky Way size. This could  help to explain the apparent early formation of super massive black holes in the Universe at $z \sim 10$. 

\acknowledgments
 We thank the anonymous referee for  useful and constructive feedback. The work of CG is supported by grants SEV-2016-0597, FPA2015-65480-P and PGC2018-095976-B-C21. The work of RJ is supported by MINECO grant PGC2018-098866-B-I00 FEDER, UE. RJ acknowledges ``Center of Excellence Maria de Maeztu 2020-2023" award to the ICCUB (CEX2019- 000918-M).

\end{document}